Influence of parasitic phases on the properties of BiFeO$_3$ epitaxial thin films

H. Béa

Unité Mixte de Physique CNRS-Thales, Domaine de Corbeville, 91404 Orsay, France

M. Bibes

Institut d'Electronique Fondamentale, Université Paris-Sud, 91405 Orsay, France

A. Barthélémy, K. Bouzehouane, E. Jacquet, A. Khodan and J.-P. Contour

Unité Mixte de Physique CNRS-Thales, Domaine de Corbeville, 91404 Orsay, France

S. Fusil

Université d'Evry, Bâtiment des Sciences, rue du père Jarlan, 91025 Evry, France

F. Wyczisk

Thales Research and Technology, Domaine de Corbeville, 91404 Orsay, France

A. Forget, D. Lebeugle, D. Colson and M. Viret

Service de Physique de l'Etat Condensé, DSM/DRECAM/SPEC, CEA Saclay, 91191 Gif-sur-Yvette, France

Abstract

We have explored the influence of deposition pressure and temperature on the growth of BiFeO$_3$ thin films by pulsed laser deposition onto (001)-oriented SrTiO$_3$ substrates. Single-phase BiFeO$_3$ films are obtained in a region close to $10^{-2}$ mbar and 580°C. In non-optimal conditions, X-ray diffraction reveals the presence of Fe oxides or of Bi$_2$O$_3$. We address the influence of these parasitic phases on the magnetic and electrical properties of the films and show that films with Fe$_2$O$_3$ systematically exhibit a ferromagnetic behaviour, while single-phase films have a low bulk-like magnetic moment. Conductive-tip atomic force microscopy mappings also indicate that Bi$_2$O$_3$ conductive outgrowths create shortcuts through the BiFeO$_3$ films, thus preventing their practical use as ferroelectric elements in functional heterostructures.

Materials that possess simultaneously several ferroic orders are called multiferroics [1]. Among these, magnetoelectrics are ferroelectric and ferro- or antiferromagnetic. The coupling between the two order parameters, called the magnetoelectric effect, is very interesting from the point of view of fundamental physics [2], and could also lead to applications in spintronics and other fields [3]. In order to use magnetoelectrics in devices, a preliminary step is to grow thin films of magnetoelectric materials. Among the possible candidates, $BiFeO_3$ (BFO) has attracted much attention [4,5,6,7]. First, it crystallises in the perovskite structure and is thus compatible with many other functional compounds. Second, it has order temperatures far beyond 300K (the ferroelectric Curie temperature is 1043K [8] and the magnetic Néel temperature 647K [9]), which is essential for applications. Third, the ferroelectric polarisation is very large, especially in thin films [7,10]. In addition, although bulk BFO is a weak ferromagnet with a magnetic moment of only ~0.01 μB/f.u. (f.u. : formula unit) [11], an increase of the magnetic moment up to about 1μB/f.u. in strained epitaxial films grown on STO(001) was recently reported [5]. This large moment may be related to a mixed $Fe^{2+}/Fe^{3+}$ valence [12] but questions still remain concerning the magnetic properties of BFO films. For the integration of BFO films into functional heterostructures, it is also crucial to investigate the stability of $BiFeO_3$, and to study the impact of possible parasitic phases on the film properties.

In this Letter, we report on the epitaxial growth of BFO films on (001) $SrTiO_3$ (STO) in a wide range of temperature $T_{dep}$ (520-750°C) and oxygen pressure P ($10^{-4}$ to $10^{-1}$ mbar). We find that the pure $BiFeO_3$ phase is formed close to P=$10^{-2}$ mbar and $T_{dep}$=580°C. At lower temperature or higher pressure, $Bi_2O_3$ precipitates are detected while at lower pressure or higher temperature, $Fe_2O_3$ forms. Magnetization measurements reveal a high magnetic moment for films containing γ-$Fe_2O_3$ impurities while single-phase films have a low magnetic moment (~0.02 μB/f.u.) close to the bulk value. The Bi-oxide impurities have the shape of square outgrowths with a large electrical conductivity that shortcut the insulating BFO film, as observed in conductive-tip atomic force microscopy (CTAFM) mappings.

The films have been grown by pulsed laser deposition on (001) STO substrates using a frequency tripled ($\lambda$=355 nm) Nd:YAG laser at a frequency of 2.5 Hz. In order to compensate for the high volatility of Bi and prevent Bi-deficiency inside the films, we have used a polycrystalline target with nominal composition $Bi_{1.15}FeO_3$. The sample temperature before and during the deposition process was measured by a double wavelength pyrometer pointing at the sample surface. The oxygen pressure during growth was varied from $10^{-4}$ to $10^{-1}$ mbar and, for each pressure, the growth rate (~0.4-0.6 $Å.s^{-1}$) was estimated from X-ray reflectometry measurements.

High-resolution X-ray diffraction (XRD) spectra (using a Panalytical X'Pert PRO equipped with a Ge(220) monochromator) for some 70 nm films grown at 580°C and different pressures are shown on figure 1a. (00$l$) peaks (labelled B) corresponding to the $BiFeO_3$ perovskite phase are visible next to the reflections of the STO substrate (labelled S). While the film grown at $6.10^{-3}$ mbar appears to be single-phase, extra peaks are detected at $10^{-4}$ mbar. Their positions match very well those of some intense reflections of $\gamma$-$Fe_2O_3$ (labelled ★), but we cannot completely exclude that these peaks may arise from $Fe_3O_4$ as these two materials have very similar crystal structures. At $10^{-1}$ mbar additional reflections corresponding to $Bi_2O_3$ (labelled ✦) are visible. Their positions correspond to the β phase (tetragonal) of this material, but could also be the signature of the δ phase (cubic) of $Bi_2O_3$. It is possible that this parasitic phase is actually a mixture of these two very similar variants.

On figure 1b, we show three θ-2θ scans for 70 nm films grown in a pressure of $1.2\ 10^{-2}$ mbar at different temperatures. The film grown at 580°C also appears to be single-phase while $Bi_2O_3$ peaks are detected at 520°C. At high temperature, the BFO phase does not form and instead (012)-oriented $\alpha$-$Fe_2O_3$ (hematite) is detected.

The results from the structural analysis of the whole films series are summarized in the pressure-temperature phase diagram of figure 2. The complexity of the diagram clearly illustrates the difficulty for optimising at the same time the Bi content integrating the film and the oxidation of Bi and Fe. As previously mentioned, no Bi-containing phase is detected at high temperature. Also, at pressures lower than $6.10^{-3}$ mbar, $\gamma$-$Fe_2O_3$ forms. The diagram shows than in these two sets of conditions (low P or high $T_{dep}$), either a large fraction of the Bi is not oxidized and then re-evaporates due to the high vapour pressure of Bi (~$10^{-3}$ mbar at 580°C) or that $Bi_2O_3$, which is unstable at high temperature [13,14], decomposes into $O_2$ and Bi which is then easily evaporated. Alternatively, at low temperature or high pressure the excess Bi present in the target oxidises in stable $Bi_2O_3$. Finally, it is only in a window close to $10^{-2}$ mbar and 580°C than single-phase films are obtained.

We now address the influence of the parasitic phases on the physical properties. As illustrated by figure 3, the magnetic properties of the films are extremely sensitive to the presence of $\gamma$-$Fe_2O_3$. Figure 3a shows a M(H) hysteresis cycle recorded at 10K for a 70 nm film grown at 580°C and $10^{-4}$ mbar, and for which a large amount of $\gamma$-$Fe_2O_3$ had been detected in X-ray diffraction. A clear ferromagnetic signal is obtained, as expected since $\gamma$-$Fe_2O_3$ is a ferrimagnet with a saturation magnetization of 480 emu.cm$^{-3}$. Even more, as the saturation magnetization for this film is about 380 emu.cm$^{-3}$, assuming the ferromagnetic signal is totally due to $\gamma$-$Fe_2O_3$, we can deduce that $\gamma$-$Fe_2O_3$ represents almost 80 % of the film volume. This value is indeed consistent with the intensity of the peaks detected in X-ray diffraction (after taking account the structure factors of the diffracting lines for BFO and $\gamma$-$Fe_2O_3$) and with compositional Auger electron spectroscopy (AES) mappings (not shown). We must mention here that very long counting rates can be necessary to detect the presence of $\gamma$-$Fe_2O_3$ in very thin (~30nm) films grown at $P \leq 2 \ 10^{-3}$ mbar, and showing a sizeable ferromagnetic signal.

In strong contrast with this observation is the very small magnetic moment detected for a single-phase film grown at 580°C at $6.10^{-3}$ mbar, see figure 3b. The magnetic properties of our single-phase films are thus consistent with those of Qi et al [15], Yun et al [16] and Eerenstein et al [17], but in strong contrast with those of Wang et al [5]. Whether it is possible to achieve large magnetic moments in single-phase films while maintaining single valence $Fe^{3+}$ ions (for example via monoclinic distortions, as sugggested by Bai et al [11]) in the BFO phase remains an open question that must be answered to define better the potential of BFO for magnetoelectric devices.

Similarly, the presence of $Bi_2O_3$ could also be detrimental for the use of BFO films in heterostructures. Indeed, although we were not able to find electrical data for $\beta$-$Bi_2O_3$, $\delta$-$Bi_2O_3$ is known as an excellent ionic conductor (with $\rho \approx 1$ $\Omega$.cm at 1073K [18]). We have observed a film containing $Bi_2O_3$ with a Scanning Auger Microscope (Nanoprobe PHI680 from Physical Electronics) equipped with a SED (Secondary Electrons Detector for classical scanning electron microscopy image) and a CMA (Cylindrical Mirror Analyser for elementary Auger mapping), in order to check its chemical homogeneity. As shown in figure 4a, the surface of this film consists of low-roughness regions and of ~100 nm-high square outgrowths. Before recording Auger spectra on this surface, a short sputtering, required to remove atmospheric contaminants (C, O), was applied. The AES composition mapping (obtained by combining elemental mappings for Bi, Fe and O) of the same area is shown in figure 4b. It indicates that the outgrowths seen by SEM are strongly Fe-deficient and correspond to a $BiO_x$ phase. We cannot quantify precisely the Bi/O ratio in these regions but they are likely to correspond to the $Bi_2O_3$ phase detected by X-ray diffraction. The regions between the outgrowths have a homogeneous (Bi, Fe, O) composition. A similar morphology, with square $\delta$-$Bi_2O_3$ hillocks has also been observed in $Bi_4Ti_3O_{12}$ films [19].

The influence of these $Bi_2O_3$ outgrowths on the electrical properties of the BFO films is exemplified in the bottom panels of figure 4. They show conductive tip AFM (CTAFM) morphology (c) and resistance (d) mappings [20] of a 70 nm thick BFO film grown on conductive Nb-doped STO, and presenting a large density of $Bi_2O_3$ outgrowths. The resistance map consists of a highly resistive background (R>$6.10^{11}$ Ω) with many low-resistance spots (R≈$10^7$ Ω). Although every outgrowth does not correspond to a conductive spot, some of them clearly create shortcuts between the metallic substrate and the conductive tip that would prevent poling the BFO in ferroelectricity measurements with micron size capacitors. Some of the $Bi_2O_3$ outgrowths therefore nucleate directly at the substrate surface. Work is in progress to understand their evolution with film thickness. In contrast, single-phase BFO films grown on Nb-doped STO (not shown) have a homogeneous resistance level that saturates the measurements capability of our system (R>$6.10^{11}$ Ω).

In summary, we have grown epitaxial BFO films onto STO(001), and explored the influence of the deposition pressure and temperature on the BFO phase purity by means of X-ray diffraction. It is found that the growth of pure BFO films is strongly favoured in a window close to 580°C and $10^{-2}$ mbar. At lower pressure or higher temperature, the presence of $Fe_2O_3$ is detected while at higher pressure or lower temperature $Bi_2O_3$ outgrowths nucleate. The presence of these two types of impurities is detrimental to the films physical properties. Indeed, films with γ-$Fe_2O_3$ show an extrinsic ferromagnetic behaviour with moments up to 2 $\mu_B$/f.u. Resistance mappings also confirm that the presence of $Bi_2O_3$ is highly problematic since $Bi_2O_3$ outgrowths create low resistance shortcuts through the BFO film. Films for which no extra phases were detected have a low, bulk-like magnetic moment and a high resistivity.

H.B. acknowledges financial support by CNRS and the Conseil Général de l'Essone.

Figure Captions :

Fig 1. (a) X-ray diffraction spectra for three films grown at 580°C and different pressures (from bottom to top, P=$10^{-4}$ mbar, 6 $10^{-3}$ mbar and $10^{-1}$ mbar). (b) Spectra for three films grown at $10^{-2}$ mbar and different temperatures (from bottom to top $T_{dep}$=520°C, 580°C and 750°C). The peaks are indexed with the following convention: B: BFO, S: STO, ★: $\gamma$-$Fe_2O_3$, O: $\alpha$-$Fe_2O_3$, ✦:$Bi_2O_3$.

Fig 2. Pressure-temperature phase diagram for BFO films with a nominal thickness of 70 nm.

Fig 3. M(H) hysteresis cycles measured at 10K with the field applied in-plane for films grown at 580°C and $10^{-4}$ mbar (a) and 580°C and 6 $10^{-3}$ mbar (b).

Fig 4. SEM image (a) and corresponding AES compositional mapping (b) for a BFO film containing $Bi_2O_3$. (b) is a RGB image constructed by superimposing coloured element-selective mappings (red : Bi, green : Fe, blue : O). The white bar scale corresponds to 1 μm. (c) and (d) are 10*10 μm² conductive tip AFM morphology (c) and resistance (d) mappings on a similar 70 nm BFO film grown on Nb-doped STO.

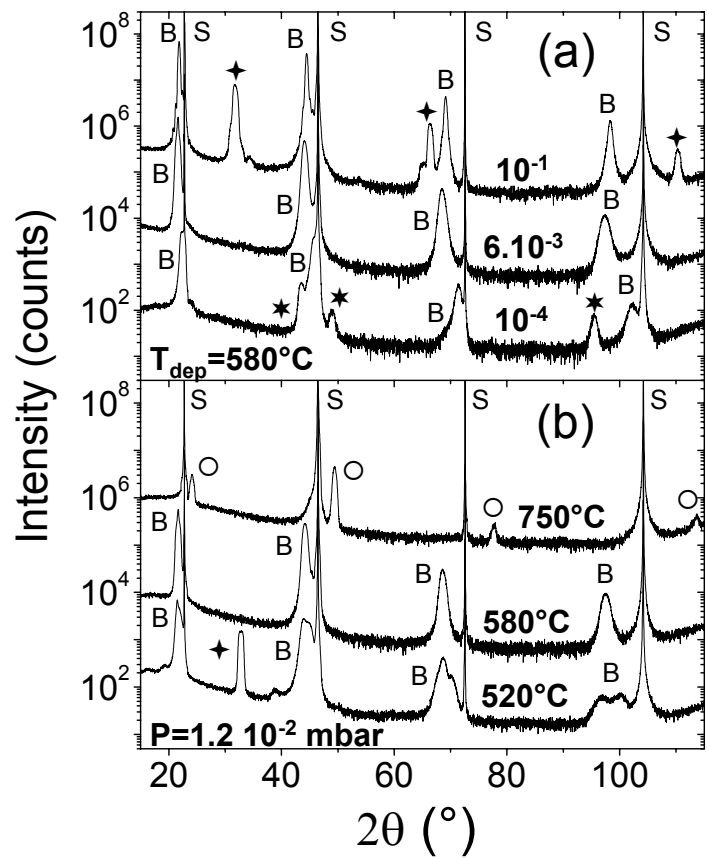

Béa et al
Fig. 1

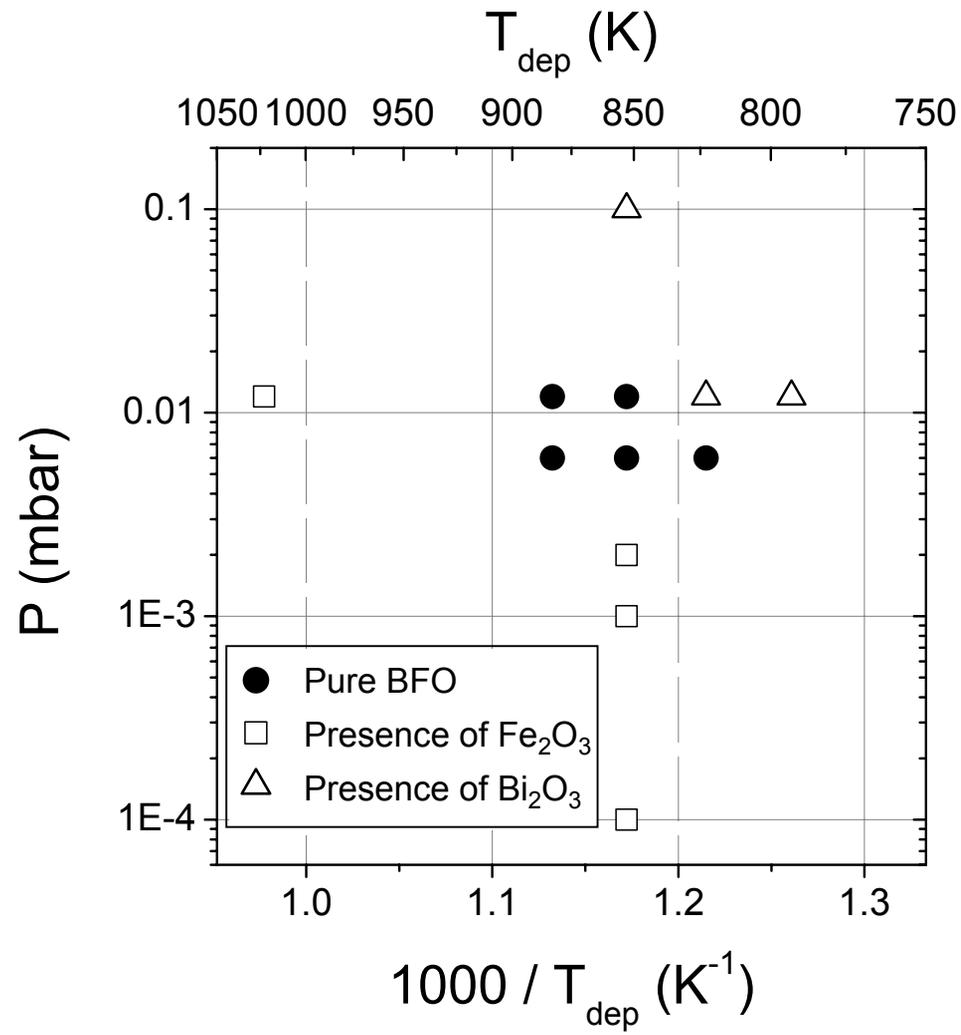

Béa et al
Fig. 2

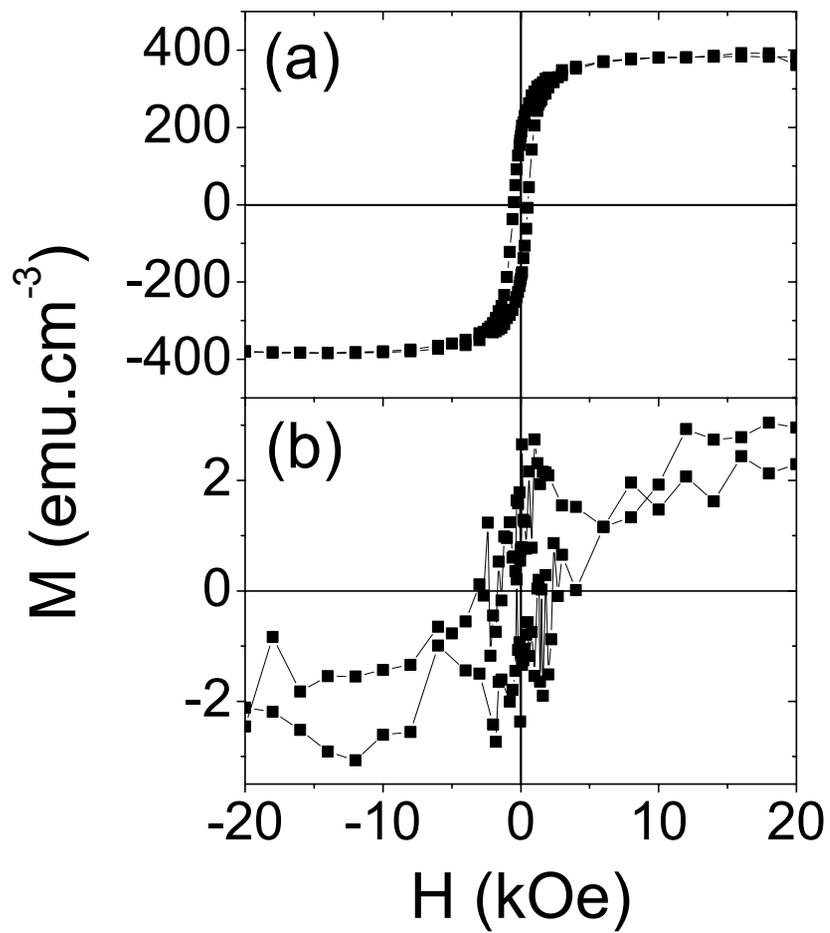

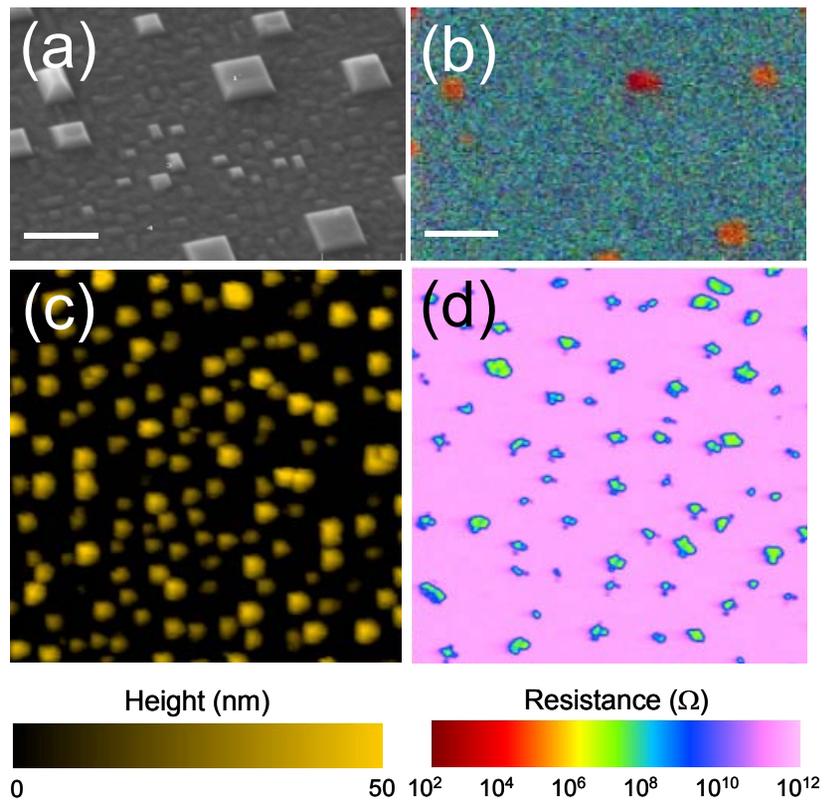

Béa et al
Fig. 4